\documentclass[reprint,twocolumn,superscriptaddress,floatfix,longbibliography]{revtex4-1}
\usepackage{latexsym}
\usepackage{lineno}
\usepackage{hyperref}
\usepackage{url}
\usepackage[caption=false,font={normalsize}]{subfig}
\usepackage{graphicx}
\usepackage{verbatim}
\usepackage{multirow}
\usepackage{amsmath, bm}
\newcommand{\beq}{\begin{eqnarray}}
\newcommand{\eeq}{\end{eqnarray}}
\usepackage{mathrsfs}
\usepackage{float}
\usepackage[usenames, dvipsnames]{color}
\usepackage{mathtools}
\usepackage{slashed}
\usepackage{physics}	
\usepackage{graphicx}   
\usepackage{hyperref}   
\usepackage{bbold}
\usepackage{amssymb}
\usepackage{feynmp-auto}
\usepackage[latin1]{inputenc}
\usepackage{tikz}
\usetikzlibrary{decorations.pathmorphing}
\usetikzlibrary{shapes.misc}
\tikzset{cross/.style={cross out, draw=black, minimum size=8*(#1-\pgflinewidth), inner sep=0pt, outer sep=0pt},
cross/.default={1pt}}
\usetikzlibrary{patterns,math}
\newcommand{\RN}[1]{%
  \textup{\uppercase\expandafter{\romannumeral#1}}%
}

\newcommand{\ncmd}{\newcommand}
\ncmd{\nn}{\nonumber}
\ncmd{\mbf}[1]{\bs{#1}}
\ncmd{\gam}{\gamma}
\ncmd{\sig}{\sigma}
\ncmd{\pha}{a}
\ncmd{\lam}{\lambda}
\ncmd{\dl}{\delta}
\ncmd{\kap}{\kappa}
\ncmd{\Lam}{\Lambda}
\ncmd{\Gam}{\Gamma}
\ncmd{\Dl}{\Delta}
\ncmd{\Ups}{\Upsilon}
\ncmd{\Om}{\Omega}
\ncmd{\eps}{\epsilon}
\ncmd{\veps}{\varepsilon}
\ncmd{\vphi}{\varphi}
\ncmd{\vtheta}{\vartheta}
\ncmd{\tw}{\text{w}}
\ncmd{\half}{\frac{1}{2}}
\ncmd{\pll}{\parallel}
\ncmd{\mc}{\mathcal}
\ncmd{\mcr}{\mathscr}
\ncmd{\mf}{\mathfrak}
\ncmd{\bs}{\boldsymbol}

\begin{document}
\title{
Shot noise in coupled electron-boson systems}
\author{Yiming Wang}
\affiliation{Department of Physics and Astronomy, Rice Center for Quantum Materials, Rice University, Houston, Texas 77005, USA}
\author{Shouvik Sur}
\affiliation{Department of Physics and Astronomy, Rice Center for Quantum Materials, Rice University, Houston, Texas 77005, USA}
\author{Chandan Setty}
\affiliation{Department of Physics and Astronomy, Rice Center for Quantum Materials, Rice University, Houston, Texas 77005, USA}
\author{Douglas Natelson}
\affiliation{Department of Physics and Astronomy, Rice Center for Quantum Materials, Rice University, Houston, Texas 77005, USA}
\author{Qimiao Si}
\affiliation{Department of Physics and Astronomy, Rice Center for Quantum Materials, Rice University, Houston, Texas 77005, USA}

\begin{abstract}
The nature of charge carriers in strange metals has become a topic of intense current investigation. Recent shot noise measurements in the 
quantum critical heavy fermion 
metal YbRh$_2$Si$_2$ revealed a suppression of the Fano factor that cannot be understood from electron-phonon scattering or strong electron correlations in a Fermi liquid, indicating loss of quasiparticles. 
The experiment motivates the consideration of shot noise in a variety of theoretical models in which quasiparticles may be lost.
Here we study shot noise in systems with co-existing itinerant electrons and dispersive bosons, going beyond the regime where the bosons are on their own
in thermal equilibrium.
We construct the Boltzmann-Langevin equations for the coupled system, and show that adequate electron-boson couplings restore the Fano factor to its Fermi liquid value. 
Our findings point to the beyond-Landau form of quantum criticality as underlying the suppressed shot noise of strange metals in heavy fermion metals and beyond.
\end{abstract}

\maketitle
\paragraph*{{\bf Introduction: }} 
In conventional metals, described by Landau Fermi liquid theory, the scattering rate increases quadratically with temperature, and the electrical current is carried by well-defined quasiparticles with electronic charge $e$~\cite{LL2013}. 
However, in strange metals, like quantum critical heavy fermion materials~\cite{Pas21.1}, resistivity  increases linearly with temperature, and quasiparticles may 
lose their identity~\cite{Hu-qcm2022, Phillips-science22}. 
This requires a new description beyond the Landau paradigm, which may no longer involve discrete current carriers. 
Shot noise provides a non-equilibrium probe of the granularity of charge carriers~\cite{Buttiker2000, Hashisaka2021}, helping to clarify the nature of current carriers in these enigmatic metals.

The shot noise Fano factor ($F$) measures the low-frequency current fluctuations relative to the average current, serving as a valuable indicator of  discreteness of the current carriers. 
Recent measurements in the quantum critical heavy fermion metal YbRh$_2$Si$_2$ revealed a strong suppression of the Fano factor in the strange metal regime~\cite{Natelson2023}, which cannot be attributed to electron-phonon interactions in a Fermi liquid. 
The experiment has motivated theoretical studies of shot noise in strongly correlated metals~\cite{Wang-Si2022,wu2023,Niko2023}.
In particular, 
we found that when the current is carried by quasiparticles, the Fano factor  
$F=\sqrt{3}/{4}$ 
is universal, being valid
even when the renormalization effect is extremely strong as in heavy Fermi liquids~\cite{Wang-Si2022}.
This implies that 
quasiparticles \textit{must} be destroyed to account for the 
shot-noise reduction, a conclusion that is consistent with the beyond-Landau form of quantum criticality advanced for 
heavy fermion metals~\cite{Si-Nature,Colemanetal,senthil2004a};
it also is supported by the overall 
phenomenology of 
YbRh$_2$Si$_2$ and related quantum critical heavy fermion metals~\cite{Wir16.1,KirchnerRMP},
which includes Fermi surface jump across the quantum critical point (QCP)~\cite{paschen2004,Friedemann.10,shishido2005}
and dynamical Planckian scaling at the QCP \cite{Schroder,Aro95.1,Prochaska2020}.
Thus,
the Fano factor joins the 
Wiedemann-Franz law~\cite{chester1961,castellani1987} 
as 
universal quantities for characterizing strong correlations in diffusive Fermi liquids and diagnosing quasiparticle loss. \par

\begin{figure}[!t]
\centering
\subfloat[\label{fig:schematic0}]{%
\includegraphics[width=0.8\columnwidth]{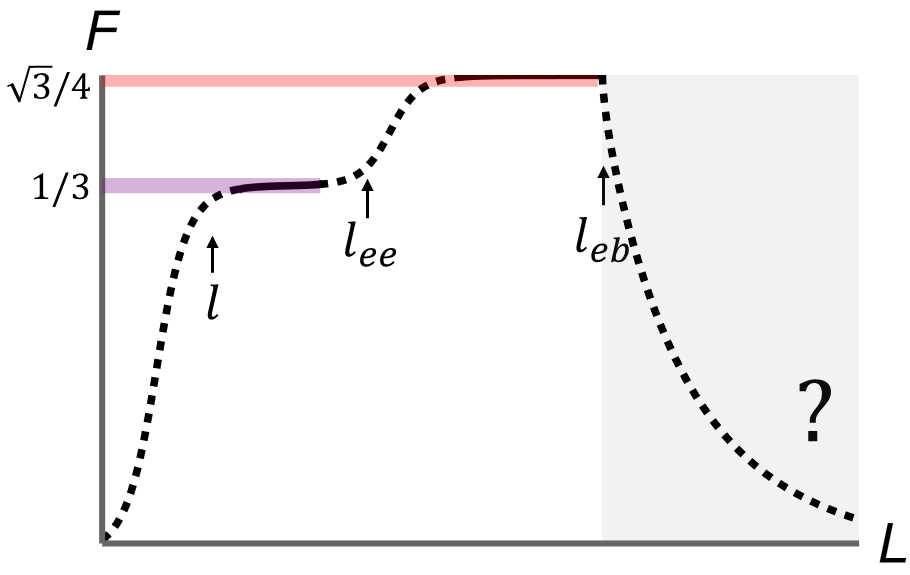}%
}
\hfill
\subfloat[\label{fig:drag}]{%
\includegraphics[width=0.85\columnwidth]{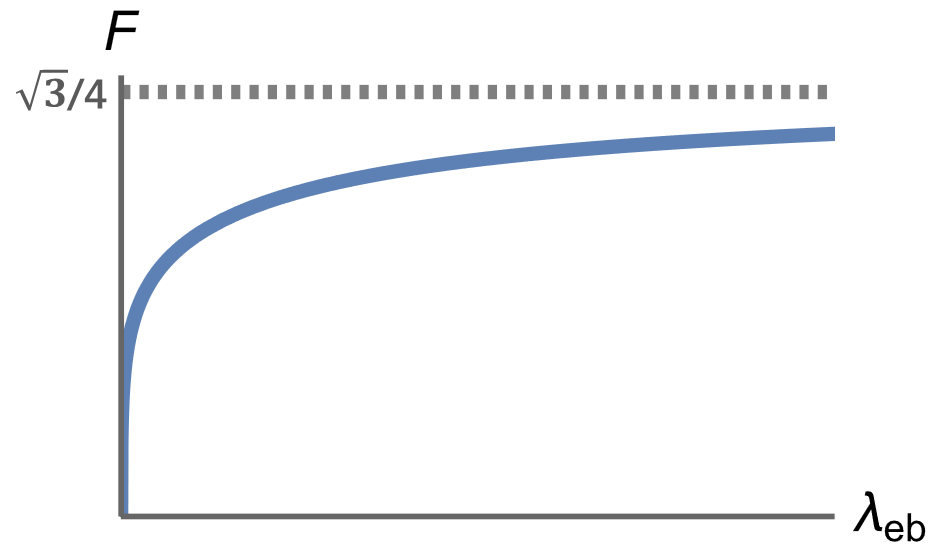}%
}
\caption{
(a) Schematic plot of the Fano factor ($F$) as a function of system size $L$,
initially derived based on a Fermi gas experiencing electron-electron and electron-phonon scattering~\cite{Devoret1996,Nagaev1995, Buttiker1992, Ueda1992}. With the phonons taken as in equilibrium on its own,
the electron-phonon coupling equilibrates the electron distribution and, thus, reduces the Fano factor, when the system size $L$ exceeds 
the relaxation lengths $l_{eb}$.
Here we ask what happens if the bosons are not assumed to be in equilibrium. (b) Schematic result for the Fano factor 
as a function of the electron-boson coupling, $\lambda_{eb}$, in the mutual drag regime (defined in the main text), when the bosons are allowed to be out of equilibrium (c.f. Fig.\ref{fig:fanodrag2}).
}
\label{fig:schematic}
\end{figure}

To further sharpen the considerations, here we study shot noise in coupled electron-boson systems. 
We take cue from the known result about the effect of electron-phonon scattering
when the phonons are in equilibrium on their own as illustrated in Fig.\,\ref{fig:schematic}(a):
when this scattering is operative, which happens when  the corresponding relaxation length $l_{eb}$ 
(over
which electrons relax to the same temperature 
as bosons) falls below
the system size $L$,
the electron-phonon coupling equilibrates the electron distribution and, thus, reduces the Fano factor~\cite{Nagaev1995, Devoret1996, Buttiker1992, Ueda1992}. 
While this phonon-based mechanism per se was ruled out for YbRh$_2$Si$_2$ through measurements of long wires~\cite{Natelson2023}, it provides a concrete setting 
to motivate
the following question: What happens to the shot noise in coupled electron-boson systems when the bosons can be driven out of equilibrium through their coupling with the electrons? 
This question is relevant to metals in the vicinity of a Landau-type quantum phase transition
where the momentum or energy transferred from the electrons to the boson can be transferred back to the electrons.
In analogy to the electron-phonon system~\cite{gurevich1989electron} a drag is present in critical metals.
In contrast to the electron-phonon system, however, this drag is mutual in critical metals because critical bosons are not independent of the underlying electrons and remain strongly coupled down to the lowest energies.
Such considerations are particularly relevant to 
the Hertz-Millis-based description of quantum criticality~\cite{lee2018recent,sur2015quasilocal,metlitski2010b}.
At density-wave criticalities,in the clean limit, only specific sections of the Fermi surface lose quasiparticle-like behavior (the so-called ``hot-spots'')~\cite{Hlubina-Rice1995, Rosch1999, schlief2017exact}  
whereas the Fermi liquid description remains valid across the remainder of the Fermi surface, dominating transport properties and leaving the Fano factor intact. 
In contrast, fluctuations of spatially uniform order parameters -- the kind relevant for ferromagnetism,
Ising-nematic
and excitonic orders
-- render the entire Fermi surface ``hot"
~\cite{lee2018recent}, 
even though their contributions to the electrical resistivity is expected to be superlinear in temperature.

In order to describe shot noise 
in such settings,
 it is desirable to construct the Boltzmann-Langevin equations suitable for 
coupled electron-boson systems without assuming equilibrium bosons.
Here, we do so 
and determine the shot noise.
Our key conclusion is that a sufficiently strong electron-boson coupling restores the Fano factor to $F=\sqrt{3}/4$, as illustrated in Fig.\,\ref{fig:schematic}(b).
Our results support the suggestion~\cite{Natelson2023,Wang-Si2022} 
that 
quantum criticality 
of beyond the Landau 
form~\cite{Si-Nature,Colemanetal,senthil2004a}, where 
criticality is driven by processes beyond a critical bosonic mode,
underlies the suppressed shot noise in YbRh$_2$Si$_2$, in a similar way that it causes a violation of the Wiedemann-Franz law~\cite{Steglich-Qi-2012}.

\paragraph*{{\bf Transport 
and fluctuations in electron-boson systems:}} Here, we consider metals with two distinct interacting degrees of freedom, one fermionic (electrons; $\psi$) and the other bosonic (phonons, collective soft modes, etc.; $\phi$), with their interaction modeled by a Yukawa term, $\phi \psi^\dagger \psi$~\footnote{Bose-Fermi systems that do not feature a  Yukawa coupling arise in ultracold atomic systems, which is beyond the scope of this paper.}
A universal description of the system in the presence of an external electromagnetic field, $\bs E$, which acts as a source terms, is given by 
\begin{align}
S = \int \dd{\tau} \dd{\bs r} \qty{\mc L[\psi, \phi]  + V(\bs r) \psi^\dagger \psi + i e \bs A(\tau, \bs r) \cdot \psi^\dagger \grad \psi},
\label{eq:action}
\end{align}
where
 $\mc L[\psi, \phi] = \psi^\dagger\qty[\partial_\tau - H_0(\grad)] \psi - \phi^\dagger \qty[\partial_\tau^2 + c^2 \nabla^2] \phi +  g(\grad)\phi \psi^\dagger  \psi + u_\psi \qty(\psi^\dagger \psi)^2 + u_\phi |\phi|^4$, $V(\bs r)$ is a random potential with $\expval{V(\bs r) V(\bs r')} \propto \delta(\bs r - \bs r')$ which is responsible for elastic scatterings among the electrons, $g(\grad)$ is a coupling function that controls the interaction between $\psi$ and $\phi$, $\bs A$ is an external electromagnetic field such that the applied electric field $\bs E = - \partial_t \bs A$. 
We will work in the regime where the self-interaction among the bosons is weak; henceforth, we set $u_\phi = 0$ and simplify notation by setting $u_\psi \to u$.
The diagrammatic representation of the remaining vertices is shown in 
Appendix C.
Here, the source of the Yukawa vertex is the four-fermion interaction channel that produces soft collective modes as the system is tuned towards a quantum critical point.
The remaining four-fermion interaction  channels are represented by the $u(\psi^\dagger \psi)^2$ term.
Thus, away from the critical point, while both interaction channels suppress the quasiparticle weight, the development of non-Fermi liquid correlations solely result from the Yukawa vertex~\cite{lee2018recent}.

\begin{figure}[!t]
\centering
\subfloat[]{
\begin{fmffile}{dia4}
\begin{fmfgraph}(80,50)
\fmfleft{i1}
\fmfright{o1}
\fmfsurroundn{v}{14}
\fmf{phantom, tension=0.8}{i1,v1}
\fmf{plain, left=0.18, tension=0.}{v2,v1}
\fmf{fermion, left=0.18, tension=0.}{v3,v2}
\fmf{fermion, tension=0.}{v4,v3}
\fmf{dots, tension=0.}{v5,v4}
\fmf{fermion, tension=0.}{v6,v5}
\fmf{fermion, left=0.18, tension=0.}{v7,v6}
\fmf{plain, left=0.18, tension=0.}{v8,v7}
\fmf{plain, left=0.18, tension=0.}{v9,v8}
\fmf{fermion, left=0.18, tension=0.}{v10,v9}
\fmf{fermion, tension=0.}{v11,v10}
\fmf{dots, tension=0.}{v12,v11}
\fmf{fermion, tension=0.}{v13,v12}
\fmf{fermion, left=0.18, tension=0.}{v14,v13}
\fmf{plain, left=0.18, tension=0.}{v1,v14}
\fmf{photon, tension=0.}{v3,v13}
\fmf{photon, tension=0.}{v6,v10}
\fmf{phantom, tension=0.8}{v6,o1}
\fmfv{decor.shape=circle,decor.filled=empty, decor.size=0.1in}{v1}
\fmfv{decor.shape=circle,decor.filled=empty, decor.size=0.1in}{v8}
\end{fmfgraph}
\end{fmffile}
}
\hfill
\subfloat[]{
\begin{fmffile}{dia5}
\begin{fmfgraph}(80,50)
\fmfleft{i1}
\fmfright{o1}
\fmfsurroundn{v}{14}
\fmf{phantom, tension=0.8}{i1,v1}
\fmf{plain, left=0.18, tension=0.}{v2,v1}
\fmf{fermion, left=0.18, tension=0.}{v3,v2}
\fmf{photon, tension=0.}{v3,v4}
\fmf{photon, tension=0.}{v4,v5}
\fmf{photon, tension=0.}{v5,v6}
\fmf{fermion, left=0.18, tension=0.}{v7,v6}
\fmf{plain, right=0.18, tension=0.}{v7,v8}
\fmf{plain, right=0.18, tension=0.}{v8,v9}
\fmf{fermion, left=0.18, tension=0.}{v10,v9}
\fmf{photon, tension=0.}{v10,v11}
\fmf{photon, tension=0.}{v11,v12}
\fmf{photon, tension=0.}{v12,v13}
\fmf{fermion, left=0.18, tension=0.}{v14,v13}
\fmf{plain, right=0.18, tension=0.}{v14,v1}
\fmf{fermion, tension=0.}{v13,v3}
\fmf{fermion, tension=0.}{v6,v10}
\fmf{phantom, tension=0.8}{v6,o1}
\fmfv{decor.shape=circle,decor.filled=empty, decor.size=0.1in}{v1}
\fmfv{decor.shape=circle,decor.filled=empty, decor.size=0.1in}{v8}
\end{fmfgraph}
\end{fmffile}
}
\hfill
\subfloat[]{
\begin{fmffile}{dia6}
\begin{fmfgraph}(60,60)
\fmfpen{thick}
\fmfleft{i1}
\fmfright{o1,o2}
\fmf{phantom}{i1,v1}
\fmf{fermion, tension=0.4}{o1,v1}
\fmf{fermion, tension=0.4}{v1,o2}
\fmfv{decor.shape=circle,decor.filled=shaded, decor.size=1em}{v1}
\end{fmfgraph}
\end{fmffile}
}
\hfill
\subfloat[]{
\begin{fmffile}{dia7}
\begin{fmfgraph}(60,60)
\fmfpen{thick}
\fmfleft{i1}
\fmfright{o1,o2}
\fmf{phantom}{i1,v1}
\fmf{boson, tension=0.4}{v1,o1}
\fmf{boson, tension=0.4}{o2,v1}
\fmfv{decor.shape=circle,decor.filled=hatched, decor.size=1em}{v1}
\end{fmfgraph}
\end{fmffile} 
}
\caption{Scattering processes contributing to the electrical conductivity. In diagrams of type (a) the external frequency-momentum can be carried entirely by electronic propagators. 
Here, one or more of the boson lines can be  exchanged for the dashed lines representing four-fermion vertices [c.f. Appendix C]. 
For diagrams of type (b), both virtual bosons and electrons carry external frequency-momentum.
These processes become important in the drag regime, as explained in the text.
(c) and (d) are the renormalized current vertices that capture the scattering of electrons and bosons by the external photon,  the thick straight (wavy) lines represent the fully renormalized electron (boson) propagators.
}
\label{fig:conductivity}
\end{figure}
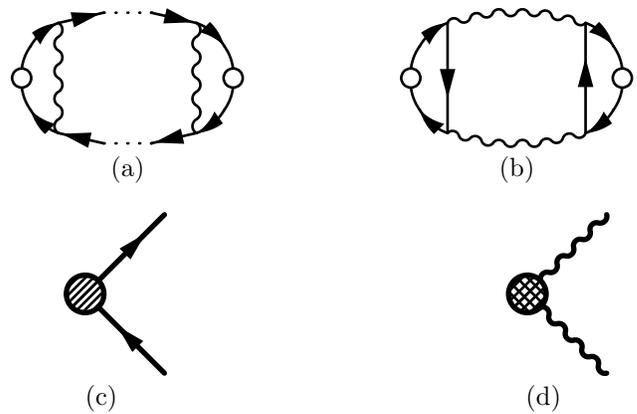

The conductivity tensor 
is given by $\sig_{ij} \sim  \expval{(\psi^\dagger \partial_i \psi) (\psi^\dagger \partial_j \psi)}$, 
which receives quantum corrections from the interaction terms in $\mc L$.
Following Holstein~\cite{Holstein1964}, we compute $\sig_{ij}$ at finite $g$ and $u$. 
Two classes of diagrams contribute to $\sig_{ij}$.
The first class of diagrams is  exemplified by processes depicted in Figs.~\ref{fig:conductivity}(a), where the external frequency-momentum can be carried entirely by electronic virtual excitations. 
The second class of diagrams, represented by Fig.~\ref{fig:conductivity}(b), constitute  scattering processes where the external frequency-momentum must  necessarily be carried by both virtual electrons and bosons.
As a result, in scattering processes of the 
latter category the bosons are 
out of equilibrium and they capture the effect of ``mutual-drag".

While computing $\sigma_{ij}$ in the mutual-drag regime, it is convenient to view the renormalizations to  
the external-source vertex (c.f., Appendix C) as a self consistent solution to a pair of coupled equations for renormalizations to the current-vertices, one for each type of current-vertex in Figs.~\ref{fig:conductivity}(c) and (d).
These equations are represented diagrammatically as 
\begin{align}
\parbox[c]{16mm}{
\begin{fmffile}{dia1a}
\begin{fmfgraph}(50,50)
\fmfpen{thick}
\fmfleft{i1}
\fmfright{o1,o2}
\fmf{phantom, tension=1}{i1,v1}
\fmf{fermion, tension=0.4}{o1,v1}
\fmf{fermion, tension=0.4}{v1,o2}
\fmfv{decor.shape=circle,decor.filled=shaded, decor.size=0.15in}{v1}
\end{fmfgraph}
\end{fmffile}
} & {}^{\displaystyle{=}}  
\parbox{16mm}{
\begin{fmffile}{dia1b}
\begin{fmfgraph}(50,50)
\fmfpen{thick}
\fmfleft{i1}
\fmfright{o1,o2}
\fmf{phantom, tension=1}{i1,v1}
\fmf{fermion, tension=0.3}{o1,v1}
\fmf{fermion, tension=0.3}{v1,o2}
\fmfv{decor.shape=circle,decor.filled=30, decor.size=0.1in}{v1}
\end{fmfgraph}
\end{fmffile}
} 
{}^{\displaystyle{+}}   
\parbox[c]{16mm}{
\begin{fmffile}{dia1c}
\begin{fmfgraph}(50,50)
\fmfpen{thick}
\fmfleft{i1}
\fmfright{o1,o2}
\fmf{phantom, tension=1}{i1,v2}
\fmf{fermion, tension=0.4}{o1,v1,v2,v3,o2}
\fmf{photon, right=0.5, tension=0}{v1,v3}
\fmfv{decor.shape=circle,decor.filled=shaded, decor.size=0.15in}{v2}
\end{fmfgraph}
\end{fmffile}
} 
{}^{\displaystyle{+}}
\parbox[c]{16mm}{
\begin{fmffile}{dia1e}
\begin{fmfgraph}(50,50)
\fmfpen{thick}
\fmfleft{i1}
\fmfright{o1,o2}
\fmf{phantom, tension=0.5}{i1,v2}
\fmf{boson, tension=0.1}{v1,v2,v3}
\fmf{fermion, tension=0.15}{o1,v1}
\fmf{fermion, tension=0.15}{v3,o2}
\fmf{fermion, tension=0}{v1,v3}
\fmfv{decor.shape=circle,decor.filled=hatched, decor.size=0.15in}{v2}
\end{fmfgraph}
\end{fmffile}
} \label{eq:coupled-1}\\ 
\parbox[c]{16mm}{
\begin{fmffile}{dia2a}
\begin{fmfgraph}(50,50)
\fmfpen{thick}
\fmfleft{i1}
\fmfright{o1,o2}
\fmf{phantom, tension=1}{i1,v1}
\fmf{photon, tension=0.4}{o1,v1}
\fmf{photon, tension=0.4}{v1,o2}
\fmfv{decor.shape=circle,decor.filled=hatched, decor.size=0.15in}{v1}
\end{fmfgraph}
\end{fmffile}
} & {}^{\displaystyle{=}}    
\parbox[c]{16mm}{
\begin{fmffile}{dia2b}
\begin{fmfgraph}(50,50)
\fmfpen{thick}
\fmfleft{i1}
\fmfright{o1,o2}
\fmf{phantom, tension=0.25}{i1,v2}
\fmf{fermion, tension=0.1}{v1,v2,v3}
\fmf{boson, tension=0.15}{o1,v1}
\fmf{boson, tension=0.15}{v3,o2}
\fmf{fermion, tension=0}{v3,v1}
\fmfv{decor.shape=circle,decor.filled=shaded, decor.size=0.15in}{v2}
\end{fmfgraph}
\end{fmffile}
}
\label{eq:coupled-2}
\end{align}
In Eq.\,\eqref{eq:coupled-1} the quantum corrections  are split into three categories: the first term is the renormalized current vertex for the electrons due to the four-fermion scatterings, the second term contains additional  corrections purely due to the Yukawa vertex, and the final term encodes the feedback from the boson-current vertex. 
Since the bosons are not charged, they do not directly couple with the external gauge field.
Consequently, the quantum corrections in Eq.\,\eqref{eq:coupled-2} are generated only through the renormalized electron-current vertex.
We note that in Eqs.~\eqref{eq:coupled-1} and \eqref{eq:coupled-2} the propagators are fully dressed by $u$, $g$, and $V$ appearing in \eqref{eq:action}
to the extent that the spectral functions still contain dispersive peaks, which implies that the fermionic excitations that contribute to the above processes are not necessarily quasiparticles. 
We 
have assumed that the system is in the regime where the Migdal's theorem applies such that the vertex corrections from the Yukawa vertex can be neglected. 

The coupled equations above can be utilized to obtain a relationship between the electron and boson distribution functions~\cite{Holstein1964}, as described in Appendix A.
In the limit $\Omega \to 0$, Eqs.~\eqref{Eq:BTE-1} and \eqref{Eq:BTE-2} lead to a set of coupled Boltzmann equations for the electrons and bosons with electron-boson ($I_{eb}$) and boson-electron ($I_{be}$) collision integtals. 
For completeness, we add collision integrals 
for the electron-electron ($I_{ee}$), boson-boson ($I_{bb}$) and electron-impurity scatterings ($I_{imp}$); 
a
Langevin source $\delta J^{ext}(x,k,t)$ 
is also present~\footnote{We note that this can be done systematically following Refs.~\cite{Wang-Si2022, betbeder1966transport}.}
This procedure leads to the following coupled Boltzmann-Langevin equations:
\begin{align}
& \hat{\mathcal{L}_\psi}f(x,k) + I_{imp}(x,k)+ I_{ee}(x,k) + I_{eb}(x,k) =\delta J^{ext} \, , \nn \\
& \hat{\mathcal{L}_\phi}N(x,q) + I_{bb}(x,q)+ I_{be}(x,q) =0 \, ,
\end{align}
where 
\begin{align}
& \hat{\mathcal{L}_\psi}f(x,\bm{k}) \nn \\
&=\qty[\partial_{t}+v_{x}\partial_{x}+(e\bm{E}-\sum_{\bm{k}'}U_{\bm{k},\bm{k}'}\partial_{x}\delta f_{\bm{k}'})\cdot\partial_{\bm{k}}]f(x,\bm{k}) \, ,\\
&    \hat{\mathcal{L}_\phi}N(x,\bm{q})=(\partial_{t}+c_{x}\partial_{x})N(x,\bm{q}) \, ,
\end{align}
and $v_{x}= \partial\epsilon_{k}/\partial k_{x}$ and $c_{x}=\partial\omega_{q}/\partial q_{x}$ are the group velocities of
the electronic and bosonic excitations, respectively, and the fermion distribution function $f$ here equals $f^+$ introduced above. 
In addition, $U_{\bm{k},\bm{k}'}$ 
describes the effective electron-electron interaction, $\delta J^{ext}$ is the Langevin source term that plays the role of an extraneous electronic flux 
for the description of 
electronic fluctuations and shot noise \cite{Nagaev1992, Wang-Si2022}.
In the correlated regime, where the electron-electron scattering length $l_{ee}$ is much smaller than the system size $L$, strong 
electron scattering significantly shapes the nonequilibrium fermion distribution. This distribution, which was derived in our previous work\cite{Wang-Si2022}, has the following form:
$f(x,k) = f_{F}(\epsilon_{\bm{k}}, T_{e}(x)) + v_{x}\tau(\partial_x +eE\partial_{\epsilon})f_{F}$, where $f_{F}(\epsilon_{\bm{k}}, T_{e})$ is the Fermi Dirac distribution function with the electron energy $\epsilon_{\bm{k}}=\epsilon_{\bm{k}}^{0}+\sum_{\bm{p}'}U_{\bm{k},\bm{k}'}\delta f_{\bm{k}'}$ and temperature $T_{e}$. The first term in this expression, symmetric in momentum space, is critical for determining the characteristics of shot noise, while the second, antisymmetric term predominantly influences the current behavior. 
We treat the electron-boson system as isolated, such that there are no additional degrees of freedom to exchange momentum and energy with the system.

\paragraph*{{\bf Shot noise:}}
When the impurity scattering is dominant over inelastic scatterings 
(i.e., at sufficiently low temperatures), the shot noise can be expressed in terms of the symmetric part of the distribution function in momentum space \cite{Wang-Si2022, Nagaev1992}: 
\begin{align}
    S=\frac{4G}{L}\int_{-L/2}^{L/2}dx\int d\epsilon f(x,\epsilon)(1-f(x,\epsilon))=4G\Bar{T}_{e} \, ,
\end{align}
where $\Bar{T}_{e}=\frac{1}{L}\int_{-L/2}^{L/2}dx T_{e}(x)$ is the averaged nonequilibrium temperature. The Fano factor, defined as the ratio between noise and current, is expressed as follows,
\begin{align}
    F=\frac{S}{2eGV}=2\Bar{\Theta}_{e} \, , 
\end{align}
where $\Bar{\Theta}_{e}\equiv \Bar{T}_{e}/eV$ denotes for the dimensionless averaged temperature. We emphasize that in the diffusive regime ($l \ll L$ where $l$ is the mean free path), the noise is predominantly determined by the symmetric part of the nonequilibrium distribution function in momemtum space, while the current, $\langle I \rangle = GV$, is solely determined by the antisymmetric part of the distribution function. 
The Fano factor is thus a result of 
both symmetric and antisymmetric contributions. 

In the absence of any electron-boson coupling,
the Fano factor $F=1/3$ 
for non-interacting electrons \cite{Nagaev1992} and $F=\sqrt{3}/4$
in the presence of 
strong electron-electron collisions \cite{Nagaev1995,Rudin1995}. 
The latter result holds true even for strongly correlated Fermi liquid \cite{Wang-Si2022}. 

The electron-boson coupling 
enables the exchange 
of energy and momentum 
between the two components.
Consequently, the temperatures of the electrons ($T_e$) and bosons ($T_b$) evolve.
It becomes necessary to solve the coupled Boltzmann-Langevin equations. 
Here, we allow the bosons  to  deviate from global equilibrium due to interactions with nonequilibrium electrons. 
Due to the locality of their collisions in real space, they tend to achieve local equilibrium at steady state. Their local equilibrium is characterized by Fermi distribution function $f_{F}(\epsilon,T_{e}(x))$ and Bose distribution function $n_B(\omega, T_b(x))$, where $T_{e}(x)$ and $T_b(x)$ denote the locally defined temperature at position $x$ within the sample. In the main text, we focus on 
critical bosons with a constant coupling function 
$g(\nabla) =g$ because the electron-boson coupling is significantly stronger in the critical boson case compared to that of Goldstone boson. 
Therefore, we need to reexamine the global equilibrium assumption for bosons, which is a prevalent approximation in electron-phonon systems.
We leave the phonon discussion into Appendix D \& E.

To investigate the evolution of $T_{e}$ and $T_{b}$, we derive the diffusion equations for electrons and bosons. They are obtained by splitting their nonequilibrium distributions into symmetric and antisymmetric part in momentum space, and substituting the Boltzmann equation for the antisymmetric part into symmetric part (see Appendix D \& E for detailed discussion):
\begin{align}
       &D\partial_{x}^{2}f(\epsilon,T_{e})=I_{eb} \, ,\label{diffe}\\
       &\partial^{2}_{x}n_{B}(\omega,T_{b})=\frac{d}{c^{2}\tau_{be}}\left(\frac{1}{\tau_{be}}+\frac{1}{\tau_{bb}}\right)[n_{B}(\omega,T_{b})-n_{B}(\omega,T_{e})] \, .\label{diffb}
   \end{align}
Here,
\begin{align}
    I_{eb}=-\int d\omega M(\omega)&[2f(\epsilon,T_{e})-f(\epsilon-\omega,T_{e})-f(\epsilon+\omega,T_{e})]\nonumber\\&[n_{B}(\omega,T_{e})-n_{B}(\omega,T_{b})] \, ,
\end{align}
where $M(\omega) =\frac{\lambda_{eb}\omega^{d-2}}{2\pi v_{F}c^{d-1}}$, $\frac{1}{\tau_{be}}=4\pi N_{F} \lambda_{eb}\frac{c}{v_{F}}$.
Here,
 $ \lambda_{eb}=g^{2}$,
 and
 $d$ is the spatial dimension. The inverse boson-boson scattering time, represented by $1/\tau_{bb}$, is assumed to be negligible relative to $1/\tau_{be}$ in this work.
$n_{B}(\omega,T_{b})$ and $n_{B}(\omega,T_{e})$ refer to Bose distribution functions of bosons and particle-hole pairs.

Since the temperatures $T_{e}$ and $T_{b}$ are the only variables to be determined in the coupled diffusion equations, we simplify them by multiplying 
Eq.\,(\ref{diffe}) with $\epsilon$ and integrate with $\epsilon$, and  multiplying 
Eq.\,(\ref{diffb}) with $\omega^{d}$ and integrate with $\omega$:
\begin{align}
&L^{2}\partial_{x}^{2}\Theta_{e}^{2}+\frac{6}{\pi^{2}}=\gamma_{e}(\Theta_{e}^{d+1}-\Theta_{b}^{d+1})\label{teme}\\
&L^{2}\partial_{x}^{2} \Theta_{b}^{d-1}=\gamma_{b}(\Theta_{b}^{d+1}-\Theta_{e}^{d+1})\label{temb}
\end{align}
where $\Theta_{e(b)}$, the dimensionless temperature for electrons (bosons),  and the parameters $\gamma_{e}$ and $\gamma_{b}$ are defined as follows 
\begin{align}
\Theta_{e(b)}= T_{e(b)}/eV; \quad 
\gamma_{e} = L^2/l_{eb}^2; \quad
\gamma_{b}= L^2/l_{b e}^2, \label{ratio}
\end{align}
with $l_{e b}^{-1} \equiv l_{e \leftarrow b}^{-1}  = \sqrt{\frac{6\zeta(d+1)\Gamma(d+1)}{2\pi^{3}v_{F}c^{d-1}D}\lambda_{eb}(eV)^{d-1}}$ and $  l_{b e}^{-1} \equiv l_{b \leftarrow e}^{-1}  = \sqrt{\frac{8d\pi^{2}N_{F}^{2}}{v_{F}^{2}}\lambda_{eb}^{2}}$. 
$\zeta(x)$ and $\Gamma(x)$ denote the Riemann zeta function and the Gamma function, respectively. $\gamma_{e(b)}$ are dimensionless quantities that characterize the energy relaxations of electrons (bosons) due to interactions with bosons (electrons), respectively. 
The terms $l_{eb}$ and $l_{be}$ denote the electron-boson and boson-electron relaxation lengths, respectively. Specifically, $l_{eb}$ measures the distance electrons can travel without energy relaxation due to bosons, and $l_{be}$ similarly applies for bosons relative to electrons. 
We note that the electron-boson relaxation length $l_{eb}$ differs from the electron-boson scattering length used in the context of resistivity,
where the latter is derived from the linearized electron-boson collision integral assuming that bosons are in equilibrium.
We further note that the ratio $\gamma_b/\gamma_e \propto \lambda_{eb}$, the electron-boson coupling at the Yukawa vertex.

Multiplying Eq.\,(\ref{temb}) with $\gamma_{e}/\gamma_{b}$ and sum with Eq.\,(\ref{teme}), one can get an exact relation between the two temperatures:
\begin{align}
    \Theta_{e}^{2}+\frac{\gamma_{e}}{\gamma_{b}}\Theta_{b}^{d+1}=\frac{6}{\pi^{2}}\left[\frac{1}{8}-\frac{1}{2}\left(\frac{x}{L}\right)^{2}\right]  \, ,\label{conserv}
\end{align}
where we utilized the zero temperature conditions at the two boundaries: $T_{e}(\pm\frac{L}{2})=T_{b}(\pm\frac{L}{2})=0$.
This relation captures the \emph{conservation law} of heat transfer between electrons and bosons under local equilibrium conditions.
Here, the second term in 
Eq.\,(\ref{conserv}) can be neglected when $\gamma_{e} \ll \gamma_{b}$, corresponding to $l_{be} \ll l_{eb}$. In this regime, the electrons stay hot, resulting in the Fano factor $F=\sqrt{3}/4$. Conversely, when $\gamma_{b} \ll \gamma_{e}$, or equivalently $l_{eb} \ll l_{be}$, the temperature of the bosons, $\Theta_{b}\rightarrow 0$. This occurs because in the absence of significant scattering from electrons ($1/\tau_{be} = 0$), bosons remain in global equilibrium.

\begin{figure}[!t]
    \centering
\includegraphics[width=\columnwidth]{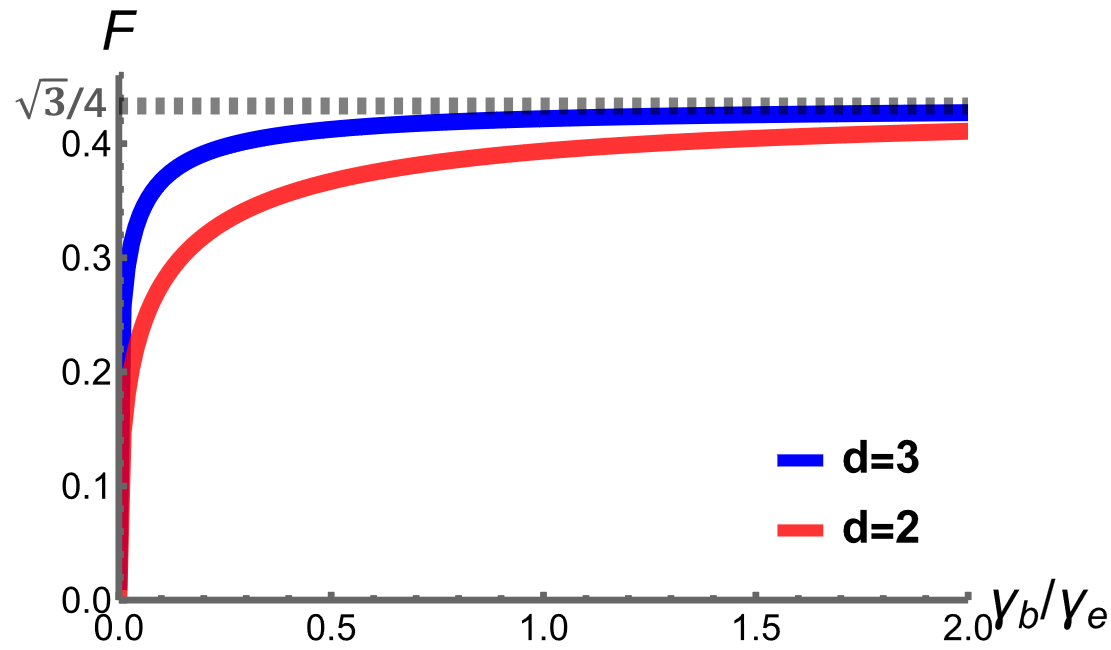}
    \caption{Fano factor $F$ as a function of the ratio $\gamma_{b}/\gamma_{e}\propto\lambda_{eb}$, in the mutual drag regime ($\gamma_{e} \gg 1$ or/and $\gamma_{b} \gg 1$.) for critical bosons, where $T_{e}(x) = T_{b}(x)$.}
    \label{fig:fanodrag2}
\end{figure}

We investigate the regime where $\gamma_{e} \gg 1$ or/and $\gamma_{b} \gg 1$, which arises 
when the electron-boson or boson-electron scattering is sufficiently strong such that 
$L \gg l_{eb}$ or $l_{be}$. The strong drag between electrons and bosons equalizes their temperatures, leading to:
\begin{align}
 \Theta_{e}=\Theta_{b} \label{drag}
\end{align}
as seen from Eqn.\,(\ref{teme} or (\ref{temb}) at leading order. 
In such a mutual-drag regime, 
we can solve analytically
for $\Theta_{e}$ after combining Eq.\,(\ref{drag}) with Eq.\,(\ref{conserv}). We summarize these results in Appendix F and plot them as a function of $\gamma_{b}/\gamma_{e}$ in Fig.~\ref{fig:fanodrag2}. 
Our analytical results are fully supported by the direct
numerical solutions to Eqs.\,(\ref{teme}, \ref{temb}).

Fig.~\ref{fig:fanodrag2} is the central result of our work. 
Note that $\gamma_b/\gamma_e \propto \lambda_{\rm eb}$, and in the mutual drag regime $L \gg l_{eb}$ or/and  $l_{be}$. 
When the electron-boson coupling $\lambda_{\rm eb}$ is small, 
$ l_{eb}\ll L, l_{be}$, and the Fano factor $F$ dips below $\sqrt{3}/4$. 
By contrast, when the electron-boson couping is adequately large such that $l_{be}\ll L,l_{eb}$, as we expect in quantum critical metals,
the Fano factor is restored to $\sqrt{3}/4$. 
This qualitative trend is illustrated in 
Fig.\,\ref{fig:schematic}(b).

\paragraph*{{\bf Discussion:}} 
Several remarks are in order. Firstly, we further stress that the mutual drag for shot noise is especially strong for the case of collective bosons. By comparison,
in an electron-phonon system the electrons couple to the displacement field, 
the Yukawa interaction is suppressed at low energies.
This feature qualitatively alters the scaling of $l_{eb}$ and $l_{be}$ with the externally applied voltage through their dependence on  $(eV)$, as detailed in Appendix E. 
An important outcome is the enhancement of the two relaxation length-scales in the electron-phonon case when compared to the critical metal. 
Thus, at weak voltages the  mutual-drag between electrons and phonons is greatly suppressed, and the two degrees of freedom equilibriate independently.
This outcome at weak $V$ is absent in critical metals, most
strikingly due to a  lack of dependence of $l_{be}$ on $(eV)$.
Therefore, the case of electrons scattering off a fluctuating, critical, local order parameter is in sharp contrast to an electron-phonon or an electron-magnon system, and the assumptions inherent in historical treatments of the latter systems do not naturally  extend to the former systems.

Secondly, we
note the striking similarity of the diffusive temperature equations (\ref{teme},\ref{temb}) with the two-temperature model in the context of time-resolved THz spectroscopy \cite{allen1987theory,barbalas2023energy}. Both reveal energy relaxation processes between electrons and bosons, suggesting the importance of enegy relaxation to the noise.
The electron-boson drag effect also occurs in the context of thermoelectricity \cite{gurevich1989electron, behnia2015fundamentals}. 
It is important to note that while both electron-boson drag in shot noise and in thermoelectricity arise from electron-boson couplings, they differ in their mechanisms. In thermoelectricity, the drag effect is due to momentum relaxation, where phonons experience the temperature gradient and feedback to the electric current
\cite{wu1996phonon, behnia2015fundamentals}. In contrast, in the shot noise context, the drag effect is primarily due to energy relaxation, where electrons and bosons exchange heat, influencing noise but not causing any significant change in the electric current.

To summarize, in this paper we study the shot noise in a coupled electron-boson system 
when the bosons are allowed to go out of equilibrium due to their interactions with the electrons.
We 
construct the coupled Boltzmann-Langevin  equations,
and show that adequate electron-boson couplings restore the Fano factor to its Fermi liquid value. 
Our results are important for understanding 
the reduced shot noise observed in quantum critical heavy fermion metals and beyond,
pointing to the quasiparticles being lost from
the beyond-Landau form of quantum criticality 
as the underlying mechanism.

\acknowledgements{
We thank 
Marco Aprilli, 
Kamran Behnia,
Matt Foster, Joel Moore,
Silke Paschen, 
Catherine P\'epin,
Sri Raghu, Subir Sachdev and Tsz Chun Wu for useful discussions.
This work has primarily been supported by the National Science Foundation under Grant No. DMR-2220603 (Y.W. and S.S.), Air Force Office of Scientific Research under
Grant No. FA9550-21-1-0356 (C.S.), 
and 
by the
Robert A. Welch Foundation Grant No. C-1411 (Q.S.) and the Vannevar Bush Faculty Fellowship ONR-VB N00014-23-1-2870 (Q.S.).
The work of D.N. was supported by the Department of Energy, Basic Energy Sciences, under Grant No. DE-FG02-06ER46337.
S.S. and Q.S. acknowledge the hospitality of the Kavli Institute
for Theoretical Physics, UCSB, supported in part by grant NSF PHY-2309135.
}

\bibliography{Noise.bib}

\clearpage
\appendix
\onecolumngrid

\section{Coupled Boltzmann equations}
The coupled equations (2,3) of the main text 
can be utilized to obtain a relationship between the electron and boson distribution functions~\cite{Holstein1964},
\begin{align} 
& i(\textbf{Q}\cdot \textbf{v}_k - \Omega) \Phi_{k} = v_k + 2\pi \sum_{\textbf q, \pm} |g_{q}|^2  \Bigg[  (\Phi_{k \pm q} -\Phi_{k} \mp n_{q} ) \left(\frac{f^{\mp}_F(\epsilon_{k\pm q}+ \Omega) + f^{\mp}_F(\epsilon_{k\pm q})}{2} + n_B(\omega_q) \right) \delta(\eps_{k}-\eps_{k+q} + \omega_q)  \nonumber \\
& \qquad + \frac{i}{2\pi} \mf{P}\left(\frac{1}{\epsilon_k - \epsilon_{k\pm q} \pm \omega_q}\right) \left( f^{\mp}_F(\epsilon_{k\pm q}+ \Omega) - f^{\mp}_F(\epsilon_{k\pm q}) \right)(\Phi_{k\pm q} - \Phi_k)\Bigg] \label{Eq:BTE-1} \\
&i(\textbf Q \cdot \nabla_q \omega_q  - \Omega ) n_{q} = 2 \pi \sum_{k}  |g_{q}|^2 \delta(\epsilon_k + \omega_q - \epsilon_{k+q})  \left[-f^-_F(\epsilon_{k}+ \omega_q) + f^-_F(\epsilon_{k}) \right]  \left[\Phi_{k+q} - \Phi_k - n_{q}\right]. \label{Eq:BTE-2}
\end{align}
Here, we have set $\hbar = 1$,  $v_k$ is the magnitude of the renormalized fermion-current vertex due to local coulomb interactions, 
$(\Omega,\textbf{Q})$ are the frequency-momentum of the external electromagnetic field, $\bs A$, $f_F^{\pm}(x)$ is the Fermi-Dirac distribution function for the occupied and vacant states, $n_B(x)$ is the Bose distribution function, $\Phi_k = \qty[f_k - f_F^-(\eps_k)]/ \qty[e E  \partial_{\epsilon_k} f_F^-(\eps_k)]$ and $n_q = \qty[N_q - n_B(\omega_q)]/\qty[e E  \partial_{\omega_q} n_B(\omega_q)]$ with $f_k$ and $N_q$ being the full fermion and boson  distribution functions, respectively, $\mf P(x)$ denotes the principal value, $g_q$ is the Fourier components of the Yukawa coupling~\footnote{For a critical boson coupled to electrons it is customary to retain the most relevant part of the electron-boson coupling function in the sense of renormalization group, and $g$ is momentum independent. By contrast, 
the phonon-electron coupling $g$ is generically momentum dependent, since phonons are Goldstone modes.}, and $\epsilon_k$ and $\omega_q$ are the
electron and boson dispersions respectively.
We note that the Thomas-Fermi screened temporal part of the electromagnetic gauge field can also participate in Boltzmann transport as an independent bosonic mode. 
Here, we assume the temperature window is sufficiently smaller than the Thomas-Fermi screening scale, such that the temporal part of the electromagnetic gauge field does not participate in transport. 
We describe the case
when the spectral functions are sharp, though we expect our analysis to remain valid provided that the spectral functions are sharp enough to display dispersive peaks.

In the limit $\Omega \to 0$, Eqs.\,\eqref{Eq:BTE-1} and \eqref{Eq:BTE-2} lead to a set of coupled Boltzmann equations for the electrons and bosons with the electron-boson ($I_{eb}$) and boson-electron ($I_{be}$) collision integtals. 
We add collision integrals resulting 
from the electron-electron ($I_{ee}$), boson-boson ($I_{bb}$), 
and electron-impurity scatterings ($I_{imp}$). Finally,
$\delta J^{ext}(x,k,t)$ represents the Langevin source~\footnote{We note that this can be done systematically following Refs.~\cite{Wang-Si2022, betbeder1966transport}. }. This leads to
the coupled Boltzmann-Langevin equations
given in Eq.\,(4) of the main text.

\section{From coupled vertex corrections to Boltzmann equations} 
In this section we outline the path from Eqs.\,\eqref{Eq:BTE-1} and \eqref{Eq:BTE-2} in the previous Appendix to the coupled Boltzmann equations. 
Here, $\Phi_k$ and $n_q$ parameterize the deviations from the equilibrium
of the fermion and boson distribution functions, respectively,
\begin{align}
f_k = f_F^-(\eps_k) + e E \Phi_k \partial_{\epsilon_k} f_F^-(\eps_k);
\qquad
N_q = n_B(\omega_q) + e E n_q \partial_{\omega_q} n_B(\omega_q).
\end{align}

In the limit $\Omega \to 0$ we obtain
\begin{align} 
& i \textbf{Q}\cdot \textbf{v}_k  \Phi_{k} = v_k + 2\pi \sum_{\textbf q, s=\pm} W_s(k,q)   (\Phi_{k +s q} -\Phi_{k} - s n_{q} ) \left(f^{-s}_F(\epsilon_{k + s q}) + n_B(\omega_q) \right) \label{eq:Phi-1} \\
&i\textbf Q \cdot \nabla_q \omega_q   n_{q} = 2 \pi \sum_{k} W_+(k,q)   \left[-f^-_F(\epsilon_{k}+ \omega_q) + f^-_F(\epsilon_{k}) \right]  \left[\Phi_{k+q} - \Phi_k - n_{q}\right] \label{eq:n-1}
\end{align}
where we have defined the transition probability
\begin{align}
W_s(k,q) = 2\pi |g_{q}|^2  \delta(\eps_{k}-\eps_{k+q} + s\omega_q)
\end{align}
We multiply both sides of 
Eq.\,\eqref{eq:Phi-1} [Eq.\,\eqref{eq:n-1}] by $e E \partial_{\eps_k} f_F(\eps_k)$ [$e E \partial_{\omega_q} n_B(\omega_q)$], and add $i \textbf{Q}\cdot \textbf{v}_k f_F$ [$i \textbf{Q}\cdot \textbf{v}_k n_B$] to obtain the Boltzmann equations.

\section{Bare vertices}
Here we specify the bare vertices associated with the interaction terms in Eq.\,(1) of the main text.
\medskip
\medskip 
\begin{figure}[!h]
\centering
\subfloat[]{
\begin{fmffile}{dia1}
\begin{fmfgraph}(60,40)
\fmfleft{i1}
\fmfright{o1,o2}
\fmf{photon, tension=0.2}{i1,v1}
\fmf{fermion, tension=0.1}{o1,v1}
\fmf{fermion, tension=0.1}{v1,o2}
\end{fmfgraph}
\end{fmffile}
}
\hfill
\subfloat[]{
\begin{fmffile}{dia2}
\begin{fmfgraph}(60,40)
\fmfleft{i1,i2}
\fmfright{o1,o2}
\fmf{fermion, tension=0.1}{i1,v1}
\fmf{fermion, tension=0.1}{v1,i2}
\fmf{dashes, tension=0.15}{v1,v2}
\fmf{fermion, tension=0.1}{o1,v2}
\fmf{fermion, tension=0.1}{v2,o2}
\end{fmfgraph}
\end{fmffile}
}
\hfill
\subfloat[]{
\begin{fmffile}{dia3}
\begin{fmfgraph}(60,40)
\fmfleft{i1}
\fmfright{o1,o2}
\fmf{phantom, tension=0.3}{i1,v1}
\fmf{fermion, tension=0.1}{o1,v1}
\fmf{fermion, tension=0.1}{v1,o2}
\fmfv{decor.shape=circle,decor.filled=empty, decor.size=0.1in}{v1}
\end{fmfgraph}
\end{fmffile}
}
\caption{Graphical representation of the three vertices that contribute to the electrical conductivity. (a) The electron-boson vertex;
(b) the four-fermion vertex; (c) the external source vertex representing the direct coupling between an applied electromagnetic field and the electron current.}
\label{fig:vertices}
\end{figure}
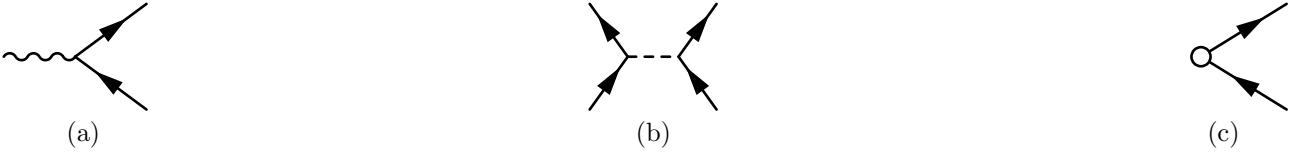

\section{Boltzmann equations for electron-boson coupled systems}
In this appendix, we start 
from the coupled Boltzmann equations,
and derive the coupled diffusion equations 
for the electrons and bosons:
\begin{align}
    \hat{\mathcal{L}}_{\psi}f(x,k) + I_{imp}(x,k)+ I_{ee}(x,k) + I_{eb}(x,k) =0 \\
    (\partial_{t}+c_{x}\partial_{x})N(x,q) + I_{bb}(x,q)+ I_{be}(x,q) =0
\end{align}
with
\begin{align}    \hat{\mathcal{L}_\psi} =\qty[\partial_{t}+v_{x}\partial_{x}+(e{E}-\sum_{{k}'}U_{{k},{k}'}\partial_{x}\delta f_{k'})\partial_{k}]
\end{align}
where the collision integrals are given by \begin{align}
    I_{imp}(x,k) = &\sum_{k'}W(kk')[f(x,k)(1-f(x,k'))-f(x,k')(1-f(x,k))], \\
    I_{ee}(x,k) = &\sum_{2,3,4}\widetilde{W}(12;34)\delta_{\bm{k}+\bm{k}_{2},\bm{k}_{3}+\bm{k}_{4}}\delta{(\epsilon+\epsilon_{2}-\epsilon_{3}-\epsilon_{4})}[f_{1}f_{2}(1-f_{3})(1-f_{4})-f_{3}f_{4}(1-f_{1})(1-f_{2})],\\  
    I_{eb}(x,k) =& 2\pi\sum_{ \bm{q}}|g_{\bm{q}}|^{2}\left\{f(x,\bm{k})[1-f(x,\bm{k}+\bm{q})]N_{\lambda}(x,\bm{q})-[1-f(x,\bm{k})]f(x,\bm{k}+\bm{q})[1+N(x,\bm{q})]\right\}\delta(\epsilon_{\bm{k}}+\omega_{\bm{q}\lambda}-\epsilon_{\bm{k+q}}) \nonumber\\
    &+|g_{\bm{q}}|^{2}\left\{f(x,\bm{k})[1-f(x,\bm{k}+\bm{q})][1+N_{\lambda}(x,\bm{q})]-[1-f(x,\bm{k})]f(x,\bm{k}+\bm{q})N(x,-\bm{q})\right\}\delta(\epsilon_{\bm{k}}-\omega_{\bm{-q}\lambda}-\epsilon_{\bm{k-q}})\\
    I_{be}(x,q) =& 4\pi\sum_{\bm{k}}|g_{\bm{q}}|^{2}\left\{N(x,\bm{q})[1-f(x,\bm{k}+\bm{q})]f(x,\bm{k})-[1+N(x,\bm{q})]f(x,\bm{k}+\bm{q})[1-f(x,\bm{k})]\right\}\delta(\epsilon_{\bm{k+q}}-\omega_{\bm{q}\lambda}-\epsilon_{\bm{k}}),
\end{align}
and $c_{x}=\frac{\partial \omega_{q}}{\partial q_{x}}=\frac{cq_{x}}{q}$ is the boson group velocity.
In these collision integrals,
    $W(pp')$ and $\widetilde{W}(12;34)$ are the scattering probability for 
    the electron-impurity and electron-electron collisions, respectively, and $|g_{q}|^{2}=\lambda_{eb}\left(\frac{\omega}{\omega_{D}}\right)^{n}$, where $n=1$ denotes for the Yukawa coupling with Goldstone bosons (e.g. phonons and AFM magnons), and $n=0$ denotes for the Yukawa coupling with critical bosons (e.g. collective soft modes of the Hubbard-Stratonovich field).

Following the procedure in the previous work \cite{Wang-Si2022}, we split the non-equilibrium Fermi distribution function into even and odd sectors:
\begin{align}
    f(x,k)= f_{even}(x,k) + f_{odd}(x,k),
\end{align}
where $f_{even}(x,k)=f_{even}(x,-k)$, $f_{odd}(x,k)=-f_{odd}(x,-k)$, and $\tau$ denotes the electron-impurity scattering time. 
The Boltzmann equation for electrons splits into two equations 
with different parities:
\begin{align}
    &(v_{x}\partial_{x}+(eE-\sum_{\bm{k}'}U_{\bm{k},\bm{k}'}\partial_{x}\delta f_{k'})\partial_{k})f_{even}=-\frac{f_{odd}}{\tau} \label{eodd}\\
    &(v_{x}\partial_{x}+(eE-\sum_{\bm{k}'}U_{\bm{k},\bm{k}'}\partial_{x}\delta f_{k'})\partial_{k})f_{odd}+I_{ee}(f_{even})+I_{eb}(f_{even})=0 \label{eeven}
\end{align}
Substituting Eqn.(\ref{eodd}) into Eqn.(\ref{eeven}), we 
have
\begin{align}
    -\tau(v_{x}\partial_{x}+(eE-\sum_{\bm{k}'}U_{\bm{k},\bm{k}'}\partial_{x}\delta f_{k'})\partial_{k})^{2}f_{even}+I_{ee}(f_{even})+I_{eb}(f_{even})=0
\end{align}
Since $f_{even}$ is an even function of $k$, one could
express $|k|$ 
through the mechanical energy $\epsilon=\epsilon_{k}^{0}+\sum_{\bm{k}'}U_{\bm{k},\bm{k}'}\delta f_{\bm{k}'}-eEx$.
Accordingly,
$\partial_{k}f_{even}(x,{k})=\frac{\partial\epsilon}{\partial k}\partial_{\epsilon}f(x,\epsilon)$, $\partial_{x}f_{even}(x,{k})=\left(\frac{\partial\epsilon}{\partial x}\partial_{\epsilon}+\partial_{x}\right)f(x,\epsilon)$.
Therefore the complex PDE is simplified with the help of the chain rule: $(v_{x}\partial_{x}+(eE-\sum_{\bm{k}'}U_{\bm{k},\bm{k}'}\partial_{x}\delta f_{k'})\partial_{k})f_{even}(x,k)=v_{x}\partial_{x}f(x,\epsilon)$, leading to the diffusion equation:

\begin{align}
    &D\partial_{x}^{2}f(x,\epsilon)=I_{ee}+I_{eb}
\end{align}
where $D=\tau v_{x}^{2}=\tau v_{F}^{2}/d$ is the diffusion constant.

When the 
electron-electron scattering is strong such that $l_{ee}\ll L$, the Fermi distribution has the following local equilibrium form 
based on the leading order equation 
of $I_{ee}(f_{even})=0$ \cite{Nagaev1995}:
\begin{align}
    f_{even}&=f_{F}(\epsilon-\mu(x),T(x))= \frac{1}{e^{(\epsilon-\mu(x))/T(x)}+1}\\
    f_{odd}&=-v_{x}\tau\partial_{x} f_{even}
\end{align}
where $\mu(x)=-eEx$ due to the boundary conditions. The full nonequilibrium distribution can be expressed as $f(x,\epsilon,v_{x})=f_{F}(\epsilon-\mu(x)-eEv_{x}\tau,T(x))$ in the weak field limit.

Next we investigate the kinetic equation for bosons. It is convenient to re-rexpress the fermion non-equilibrium fermion distribution in the $k$ basis, 
\begin{align}
    f(x,k)= f_{F}(\epsilon_{\bm{k}}-eEv_{x}\tau, T_{e}(x)) \label{fermi}
\end{align}
The form of $I_{be}$ can be simplified with the help of 
Eq.\,{\ref{fermi}}, 
\begin{align}
    I_{be}(x,q)=&-4\pi|g_{q}|^{2}[n_{B}(\omega_{q},T_{e}(x))-N(x,q)]\sum_{k}[f(\epsilon_{k}-eEv_{F}\tau)-f(\epsilon_{k}+\omega_{q}-eEv_{F}\tau)]\delta(\epsilon_{\bm{k+q}}-\omega_{\bm{q}\lambda}-\epsilon_{\bm{k}})\\
    =&-4\pi N_{F}|g_{q}|^{2}\frac{c}{v_{F}}[n_{B}(\omega_{q},T_{e})-N(x,q)]
\end{align}
where 
\begin{align}
    \sum_{k} [f(\epsilon_{k})-f(\epsilon_{k+q})]\delta(\omega-\epsilon_{k+q}+\epsilon_{k})=-\frac{1}{\pi}Im[\sum_{k}\frac{1}{\omega-\epsilon_{k+q}+\epsilon_{k}+i\delta}]=N_{F}\frac{\omega}{v_{F}q}
\end{align}
The nonequilibrium boson distribution could also be splitted into even and odd sectors:
\begin{align}
    N(x,q)= N_{even}(x,q)+N_{odd}(x,q)
\end{align}
where $N_{even}=n_{B}(\omega_{q},T_{e})$ preserves under $q\rightarrow -q$, and $N_{odd}(x,q)= -N_{odd}(x,-q)$.
Due to different parities, the boson Boltzmann equation is split into two equations:
\begin{align}
    c_{x}\partial_{x}N_{odd} &= [n_{B}(\omega_{q},T_{e})-N_{even}(x,q)]/\tau_{be}\label{pheven}\\
    c_{x}\partial_{x}N_{even} &= -N_{odd}(x,q)(\frac{1}{\tau_{be}(q)}+\frac{1}{\tau_{bb}(q)}) \label{phodd}
\end{align}
with $1/\tau_{be}=4\pi N_{F} |g_{q}|^{2}\frac{c}{v_{F}}=4\pi N_{F} \lambda_{eb}\frac{c\omega^{n}}{v_{F}\omega_{D}^{n}}$. Combining Eqs.(\ref{pheven},\ref{phodd}), we have
\begin{align}
    c_{x}^{2}\partial_{x}^{2}n_{B}(\omega_{q},T_{b}(x)) =\frac{1}{\tau_{be}}\left(\frac{1}{\tau_{be}}+\frac{1}{\tau_{bb}}\right)[n_{B}(\omega_{q},T_{b}(x))-n_{B}(\omega_{q},T_{e}(x))]
\end{align}
where $c_{x}=\frac{\partial \omega_{q}}{\partial q_{x}}=\frac{cq_{x}}{q}$, $c^{2}_{x}=\frac{1}{d}c^{2}$.

On the other hand, from the electron side, the collision integral can be simplified at leading order (we do not consider Umklapp scattering: $\bm{G} = 0$),
\begin{align}
    I_{eb}(x,k) =& 2\pi\sum_{ \bm{q}}|g_{\bm{q}}|^{2}\left\{f(x,\bm{k})[1-f(x,\bm{k}+\bm{q})]N(x,\bm{q})-[1-f(x,\bm{k})]f(x,\bm{k}+\bm{q})[1+N(x,\bm{q})]\right\}\delta(\epsilon_{\bm{k}}+\omega_{\bm{q}\lambda}-\epsilon_{\bm{k+q}}) \nonumber\\
    &+|g_{\bm{q}}|^{2}\left\{f(x,\bm{k})[1-f(x,\bm{k}+\bm{q})][1+N(x,\bm{q})]-[1-f(x,\bm{k})]f(x,\bm{k}+\bm{q})N(x,-\bm{q})\right\}\delta(\epsilon_{\bm{k}}-\omega_{\bm{-q}\lambda}-\epsilon_{\bm{k-q}})\\
    =&\int d\omega M(\omega)\left\{f(\epsilon_{\bm{k}},T_{e}))[1-f(\epsilon_{\bm{k}}+\omega, T_{e})]n_{B}(\omega, T_{b})-[1-f(\epsilon_{\bm{k}},T_{e})]f(\epsilon_{\bm{k}}+\omega, T_{e})[1+n_{B}(\omega,T_{b})] \right.\nonumber\\ &\left.-[1-f(\epsilon_{\bm{k}},T_{e})]f(\epsilon_{\bm{k}}-\omega,T_{e})N(\omega, T_{b})+f(\epsilon_{\bm{k}},T_{e})[1-f(\epsilon_{\bm{k}}-\omega,T_{e})][1+N(\omega,T_{b})]\right\}\\
    =&-\int d\omega M(\omega) [2f(\epsilon_{\bm{k}},T_{e})-f(\epsilon_{\bm{k}}-\omega,T_{e})-f(\epsilon_{\bm{k}}+\omega,T_{e})][n_{B}(\omega,T_{e})-n_{B}(\omega,T_{b})],
\end{align}
where
\begin{align}
    M(\omega)= 2\pi\int\frac{d^{d}\bm{q}}{(2\pi)^{d}}|g_{q}|^{2}\delta(\omega-\omega_{q})\delta(\epsilon_{k}+\omega_{q}-\epsilon_{k+q}).
\end{align}
For parabolic electronic bands we have,
\begin{align}
    \omega_{q}+\epsilon_{k}-\epsilon_{k+q}=(c-v_{F}\cos\theta)q + \frac{q^{2}}{2m}.
\end{align}
The form of $M(\omega)$ is dimensionality dependent. For small $\bm{q}$ in three dimensions
\begin{align}
    M(\omega) = \frac{1}{2\pi}\frac{\lambda}{v_{F}c^{2}\omega_{D}^{n}} \omega^{n+1},
\end{align}
while in two dimensions
\begin{align}
    M(\omega)= \frac{1}{2\pi}\frac{\lambda \omega}{v_{F}c\sin(\arccos{(c/v_{F})})} \approx\frac{1}{2\pi}\frac{\lambda \omega^{n}}{v_{F}c\omega_{D}^{n}}.
\end{align}

After combining the even and odd sectors of the Boltzmann equations \cite{Wang-Si2022}, we 
derive the following coupled electron-boson diffusion equations:
   \begin{align}
       &D\partial_{x}^{2}f(\epsilon,T_{e})=I_{eb}\\
       &\partial^{2}_{x}n_{B}(\omega,T_{b})=d(4\pi N_{F}\lambda/v_{F}\omega_{D}^{n})^{2}\omega^{2n}[n_{B}(\omega,T_{b}(x))-n_{B}(\omega,T_{e}(x))].
   \end{align}
where 
\begin{align}
    I_{eb}=-\int d\omega M(\omega)&[2f(\epsilon,T_{e})-f(\epsilon-\omega,T_{e})-f(\epsilon+\omega,T_{e})]\nonumber\\&[n_{B}(\omega,T_{e})-n_{B}(\omega,T_{b})]
\end{align}
where $M(\omega) =\frac{\lambda_{eb}\omega^{n+d-2}}{2\pi v_{F}\omega_{D}^{n}c^{d-1}}$, $\frac{1}{\tau_{be}}=4\pi N_{F} \lambda_{eb}\frac{c}{v_{F}}\left(\frac{\omega}{\omega_{D}}\right)^{n}$. Here, $d$ is the spatial dimension, $n=1$ describes the momentum-scaling of the Yukawa coupling
in the case of Goldstone bosons (e.g. phonons and AFM magnons), 
whereas $n=0$ corresponds to the Yukawa coupling 
for critical bosons (e.g. collective soft modes of the Hubbard-Stratonovich field), which correspond to 
Eqs.\,(\ref{diffe} and \ref{diffb}) in the main text.

\section{Analysis of the coupled diffusion equations}

Using the coupled diffusion equations 
derived in the previous section, we 
investigate the temperature profiles.

Multiplying Eq.\,(\ref{diffe})
by $\epsilon$ and 
integrating with $\epsilon$, 
and, in addition, 
multiplying 
Eq.\,(\ref{diffb}) 
by $\omega^{d-n}$ and integrating
with $\omega$,
one gets the temperature equations 
that correspond to 
Eqs.\,(\ref{teme},\ref{temb}) for critical bosons ($n=0$) in the main text:
\begin{align}
&L^{2}\partial_{x}^{2}\Theta_{e}^{2}+\frac{6}{\pi^{2}}=\gamma_{e}(\Theta_{e}^{d+n+1}-\Theta_{b}^{d+n+1})\label{eteme}\\
&L^{2}\partial_{x}^{2} \Theta_{b}^{d-n+1}=\gamma_{b}(\Theta_{b}^{d+n+1}-\Theta_{e}^{d+n+1})\label{etemb}
\end{align}
where $\Theta_{e(b)}$, the dimensionless temperature for electrons (bosons),  and the parameters $\gamma_{e}$ and $\gamma_{b}$ are defined as follows: 
\begin{align}
\Theta_{e(b)}= T_{e(b)}/eV; \quad 
\gamma_{e} = L^2/l_{eb}^2; \quad
\gamma_{b}= L^2/l_{b e}^2, 
\end{align}
with $l_{e b}^{-1} \equiv l_{e \leftarrow b}^{-1}  = \sqrt{\frac{6\zeta(d+n+1)\Gamma(d+n+1)}{2\pi^{3}v_{F}c^{d-1}\omega_{D}^{n}D}\lambda_{eb}(eV)^{d+n-1}}$ and $  l_{b e}^{-1} \equiv l_{b \leftarrow e}^{-1}  = \sqrt{\frac{8d\pi^{2}N_{F}^{2}}{v_{F}^{2}\omega_{D}^{2n}}\frac{\zeta(d+n+1)\Gamma(d+n+1)}{\zeta(d-n+1)\Gamma(d-n+1)}\lambda_{eb}^{2}(eV)^{2n}}$. 
Here,
$\zeta(x)$ and $\Gamma(x)$ denote the Riemann zeta function and the Gamma function, respectively,
and $\gamma_{e(b)}$ are dimensionless quantities that characterize the energy relaxations of electrons (bosons) due to interactions with bosons (electrons), respectively. 
The terms $l_{eb}$ and $l_{be}$ denote the electron-boson and boson-electron relaxation lengths, respectively. Specifically, $l_{eb}$ measures the distance electrons can travel without energy relaxation due to bosons, and $l_{be}$ similarly applies for bosons relative to electrons. 
We note that the electron-boson relaxation length $l_{eb}$ differs from the electron-boson scattering length used in the context of resistivity (e.g., Bloch's law for acoustic phonons), where the latter is derived from the linearized electron-boson collision integral assuming that bosons are in equilibrium.
We further note that the ratio $\gamma_b/\gamma_e \propto \lambda_{eb}$, the electron-boson coupling at the Yukawa vertex.

Multiplying Eq.\,(\ref{temb}) 
by $\gamma_{e}/\gamma_{b}$ and 
summing with Eq.\,(\ref{teme}), one can get an exact relation between the two temperatures:
\begin{align}
    \Theta_{e}^{2}+\frac{\gamma_{e}}{\gamma_{b}}\Theta_{b}^{d-n+1}=\frac{6}{\pi^{2}}\left[\frac{1}{8}-\frac{1}{2}\left(\frac{x}{L}\right)^{2}\right] \label{econserv}
\end{align}
where we utilized the zero temperature conditions at the two boundaries: $T_{e}(\pm\frac{L}{2})=T_{b}(\pm\frac{L}{2})=0$.
This relation captures the \emph{conservation law} of heat transfer between the electrons and bosons under local equilibrium conditions.
Here, the second term in 
Eq.\,(\ref{econserv}) can be neglected when $\gamma_{e} \ll \gamma_{b}$, corresponding to $l_{be} \ll l_{eb}$. In this regime, the electrons stay hot, resulting in Fano factor $F=\sqrt{3}/4$. Conversely, when $\gamma_{b} \ll \gamma_{e}$, or equivalently $l_{eb} \ll l_{be}$, the temperature of the bosons, $\Theta_{b}\rightarrow 0$. This occurs because in the absence of significant scattering from electrons ($1/\tau_{be} = 0$), bosons remain in global equilibrium.

We investigate the regime where $\gamma_{e} \gg 1$ or $\gamma_{b} \gg 1$, which arises 
when the electron-boson or boson-electron scattering is sufficiently strong such that 
$L \gg l_{eb}$ or $l_{be}$. The strong drag between 
the electrons and bosons equalizes their temperatures:
\begin{align}
 \Theta_{e}=\Theta_{b} \, ,\label{edrag}
\end{align}
as seen from Eq.\,(\ref{eteme} or (\ref{etemb}) at the leading order.
Here,
one can derive 
analytic solutions 
for $\Theta_{e}$ after combining 
Eq.\,(\ref{edrag}) with Eq.\,(\ref{econserv}).

\section{Analytic expression of $\Theta_{e}$ in the 
mutual-drag regime}

In the electron-boson 
mutual-drag
regime, the bosons and electrons share the same temperatures: $\Theta_{e}=\Theta_{b}$. After combining with 
Eq.\,(\ref{conserv}) and 
Eq.\,(\ref{drag}) in the main text,
\begin{align}
    \Theta_{e}&=\Theta_{b}\\
    \Theta_{e}^{2}+\frac{\gamma_{e}}{\gamma_{b}}\Theta_{b}^{d-n+1}&=\frac{6}{\pi^{2}}\left[\frac{1}{8}-\frac{1}{2}\left(\frac{x}{L}\right)^{2}\right]
\end{align}
one can solve $\Theta_{e}$, 
for
different dimension $d$ and
different exponent $n$ of the momentum dependence in the
Yukawa coupling, as functions of $p=\gamma_{b}/\gamma_{e}$ and $\Theta_{0e}(x)=\sqrt{\frac{3}{4\pi^{2}}\left[1-\left(\frac{2x}{L}\right)^{2}\right]}$:

\begin{itemize}
    \item $d=2, n=1$:
        \begin{align}
            \Theta_{e}(p,\Theta_{0e})
            =&\sqrt{\frac{p}{1+p}}\Theta_{0e}
        \end{align}
    \item $d=3, n=0$:
          \begin{align}  \Theta_{e}(p,\Theta_{0e})=\frac{1}{\sqrt{2}}\sqrt{\sqrt{p^{2}+4p\Theta_{0e}^{2}}-p}
        \end{align}
    \item $d=2, n=0$ and $d=3, n=1$: 
        \begin{align}
            \Theta_{e}(p,\Theta_{0e})=\frac{1}{3} \left(-p + \frac{2^{1/3} p^2}{\left(27 \Theta_{0e}^2 p - 2 p^3 + 3 \sqrt{81 \Theta_{0e}^4 p^2 - 12 \Theta_{0e}^2 p^4}\right)^{1/3}} + \frac{\left(27 \Theta_{0e}^2 p - 2 p^3 + 3 \sqrt{81 \Theta_{0e}^4 p^2 - 12 \Theta_{0e}^2 p^4}\right)^{1/3}}{2^{1/3}}\right)\label{expression2}
        \end{align} 
\end{itemize}
We also summarize these results in 
Table\,I.
\begin{table}[!t]
\centering
\begin{tabular}{|c|c|c|}
\hline
 $\Theta_{e}^{(d-n)}(x)$& \textbf{$n = 0$} & \textbf{$n = 1$} \\
\hline
\textbf{$d = 2$} & $\Theta^{(2)}\left(\frac{\gamma_{b}}{\gamma_{e}},\Theta_{0e}\right)$ & $\sqrt{\frac{\gamma_{b}}{\gamma_{e}+\gamma_{b}}}\Theta_{0e}$ \\
\hline
\textbf{$d = 3$} & $\frac{1}{\sqrt{2}}\left( \sqrt{\frac{\gamma_{b}^{2}}{\gamma_{e}^{2}}+\frac{4\gamma_{b}}{\gamma_{e}}\Theta_{0e}^{2}}-\frac{\gamma_{b}}{\gamma_{e}}\right)^{\frac{1}{2}}$ & $\Theta^{(2)}\left(\frac{\gamma_{b}}{\gamma_{e}},\Theta_{0e}\right)$ \\
\hline
\end{tabular}
\caption{\label{table:temx}
Local temperature for electrons, 
in the cases of electron-coupling to
critical $(n=0)$ and Goldstone $(n=1)$ bosons in two ($d=2$) and three ($d=3$) dimensions, in the limit $\gamma_{e}$ or $\gamma_{b}\gg 1$. $\Theta_{0e}(x)=\sqrt{\frac{3}{4\pi^{2}}\left[1-\left(\frac{2x}{L}\right)^{2}\right]}$ is the local temperature of hot electrons without electron-boson couplings, which leads to $F(\lambda_{eb}=0)=\frac{\sqrt{3}}{4}$. The form of $\Theta^{(2)}$ is shown in Eq.(\ref{expression2}).}
\end{table}

\twocolumngrid
\end{document}